# Security Issues on Cloud Computing

Harit Shah
Hshah10@iit.edu

Sharma Shankar Anandane
sharmapdy06@gmail.com

Shrikanth
ssundar8@gmail.com

***Abstract***

**The Cloud Computing concept offers dynamically scalable resources provisioned as a service over the Internet. Economic benefits are the main driver for the Cloud, since it promises the reduction of capital expenditure and operational expenditure. In order for this to become reality, however, there are still some challenges to be solved. Amongst these are security and trust issues, since the user's data has to be released to the Cloud and thus leaves the protection sphere of the data owner. Most of the discussions on these topics are mainly driven by arguments related to organisational means. This paper focuses on various security issues arising from the usage of Cloud services and especially by the rapid development of Cloud computing arena. It also discusses basic security model followed by various High Level Security threats in the industry.**

## I. INTRODUCTION

Cloud computing is a model for enabling convenient, on-demand network access to a shared pool of configurable computing resources (e.g., networks, servers, storage, applications, and services) that can be rapidly provisioned and released with minimal management effort or service provider interaction. Cloud computing architecture, just like any other system, is categorized into two main sections: Front End and Back End. Front End can be end user or client or any application (i.e. web browser etc.) which is using cloud services. Back End is the network of servers with any computer program and data storage system. It is usually assumed that cloud contains infinite storage capacity for any software available in market. Cloud has different applications that are hosted on their own dedicated server farms. Cloud has centralized server administration system. Centralized server administers the system, balances client supply, adjusts demands, monitors traffic and avoids congestion. This server follows protocols, commonly known as middleware. Middleware controls the communication of cloud network among them. Cloud Architecture runs on a very important assumption, which is mostly true. The assumption is that the demand for resources is not always consistent from client to cloud. Because of this reason the servers of cloud are unable to run at their full capacity. To avoid this scenario, server virtualization technique is applied. In sever virtualization, all physical servers are virtualized and they run multiple servers with either same or different application. As one physical server acts as multiple physical servers, it curtails the need for more physical machines. As a matter of fact, data is the most important part of cloud computing; thus, data security is the top most priority in all the data operations of cloud. Here, all the data are backed up at multiple locations. This astoundingly increases the data storage to multiple times in cloud compared with a regular system. Redundancy of data is crucial, which is a must-have attribute of cloud computing.

Security of confidential data (e.g., SSN or Credit Card Numbers) is a very important area of concern as it can make way for very big problems if unauthorized users get access to it. Misuse of data can create big issues; hence, in cloud computing it is very important to be aware of data administrators and their extent of data access rights. Large organizations dealing with sensitive data often have well laid out regulatory compliance policies. However, these polices should be verified prior to engaging them in cloud computing. There is a possibility that in cloud computing network, sometimes the network utilizes resources from another country or they might not be fully protected; hence, the need arises for appropriate regulatory compliance policies. In cloud computing, it is very common to store data of multiple customers at one common location. Cloud computing should have proper techniques where data is segregated properly for data security and confidentiality. Care must be taken to ensure that one customer's data does not affect another customer's data. In addition, Cloud computing providers must be equipped with proper disaster recovery policies to deal with any unfortunate event.

If we see architecture of the cloud computing that is used in many areas of current research and corporate world then it has Security as an important factor for selection of cloud configuration. Fig 1 shows the importance of security for various types of Clouds that are existing in current world.

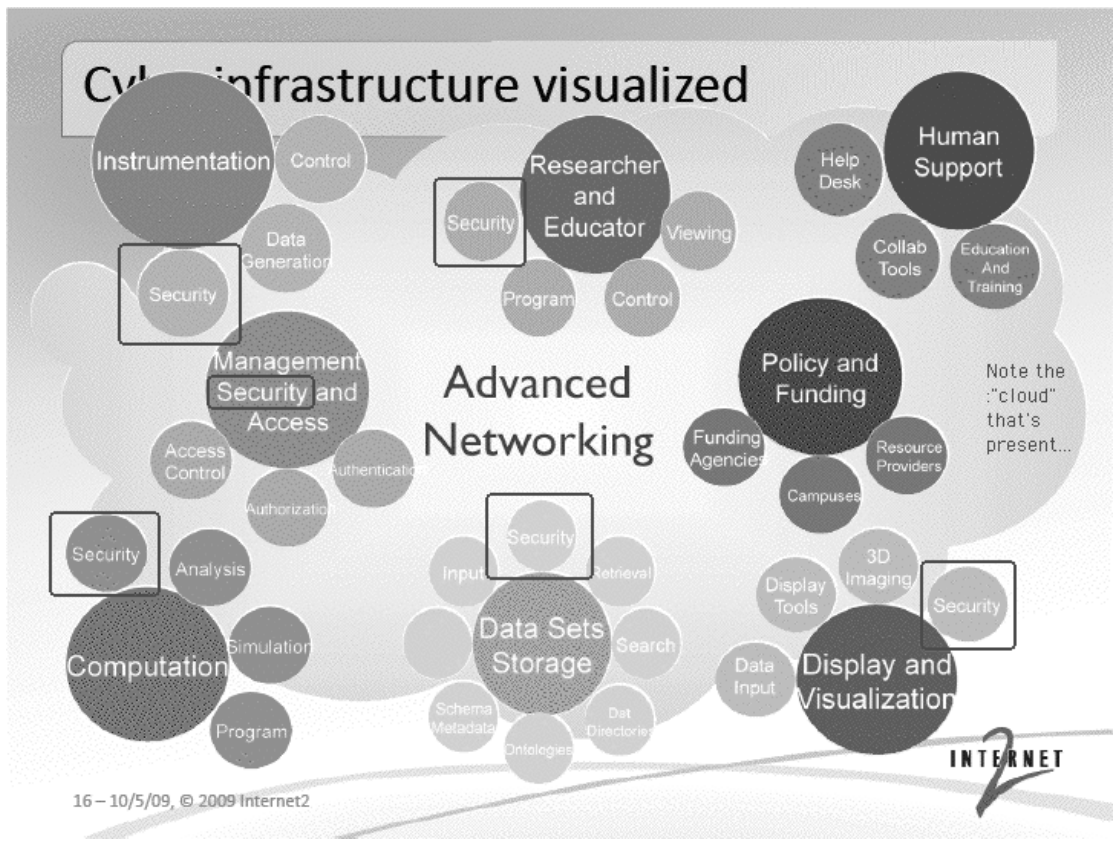

[Fig. 1]

According to the survey carried by NIST [1], for most of the big companies security is biggest concern for migrating their product to cloud. Cloud computing has lucrative offers economically and on the technical part but they are still concerned about the security managed by cloud they will hire. Fig-2 shows the statistics of the survey.

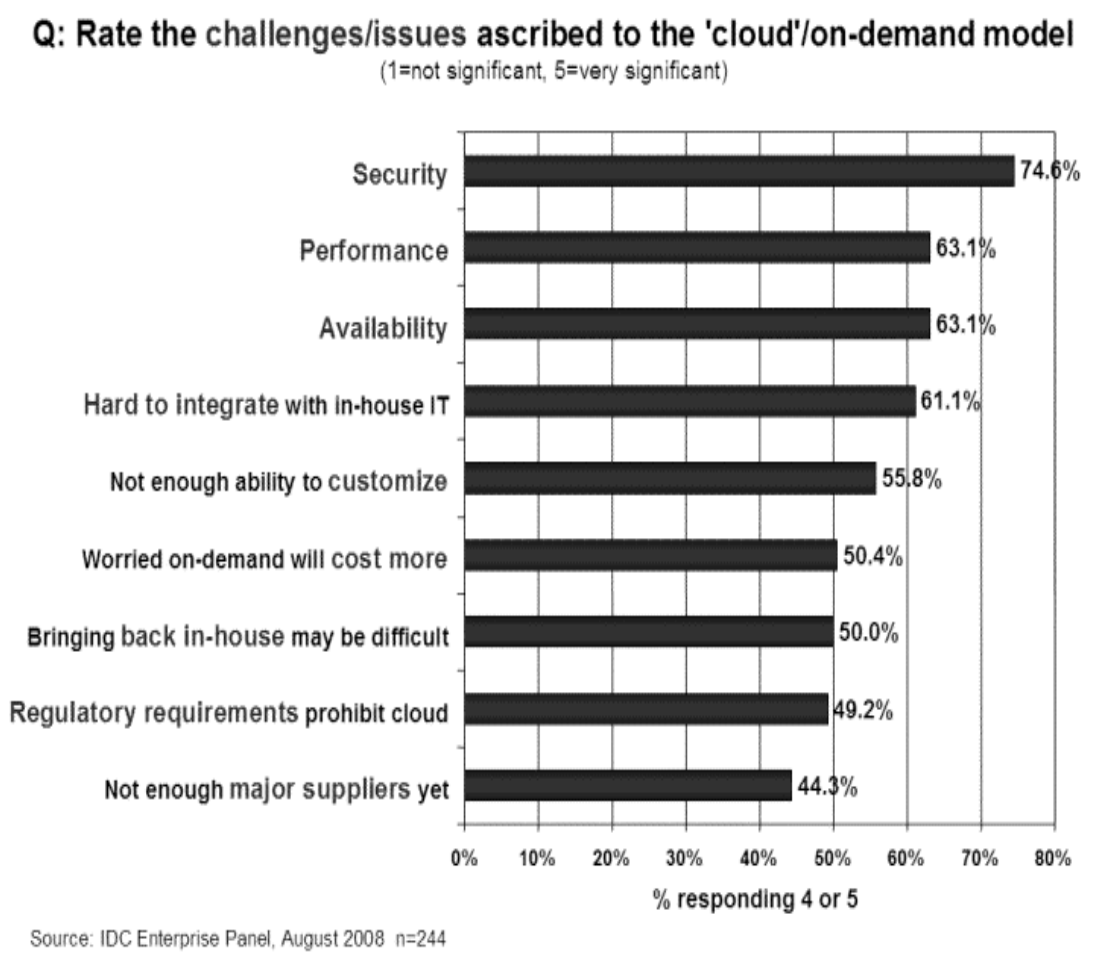

[Fig-2]

In this paper we will first discuss various aspects of security with security model that is been proposed by Jericho Forum[2], Followed by that we have discussed the CIA objectives of security related to cloud computing. In section 4 we have discussed major threats in current world by categorizing them in Computation Security, Storage Security and Network Security. Each sub-section discusses the priority, reasons for those threats, and repercussions of that threat and possible solutions that are currently accepted by the industry. In the end paper is concluding about the current severity on security issues in cloud computing.

## II. Security Analysis

Basically Cloud model can be broken down in mainly three layers: 1. Infrastructure as a service (IaaS) 2. Platform as a Service (PaaS) and 3. Software as a Service (SaaS). Here security for each layer has different issues but still they can be closely combined in to one cardinal framework. Jericho Forum has proposed a model for cloud computing which integrates Security (and Identity Managers) inside the layers of the cloud computing. Fig-3 shows the pictorial view of the Cloud Computing model. For evaluating the security for any cloud there are mainly CIA objectives are to be taken in consideration. CIA analysis includes 1. Confidentiality 2. Integrity and 3. Availability. For anyone to select the cloud provider one must have to consider the CIA objectives. **Confidentiality** is one of the prime constraints for the growth of cloud computing paradigm. Users when select the Cloud provide they must be sure that the data that is given to the provider must be confidential. Provider must protect it from other users as well as must provide surety that even provider will also not peep into the data. Typically confidentiality is maintained by the encryption of the data that has been uploaded on the server of provider. But encryption has huge drawback in performance of the system.

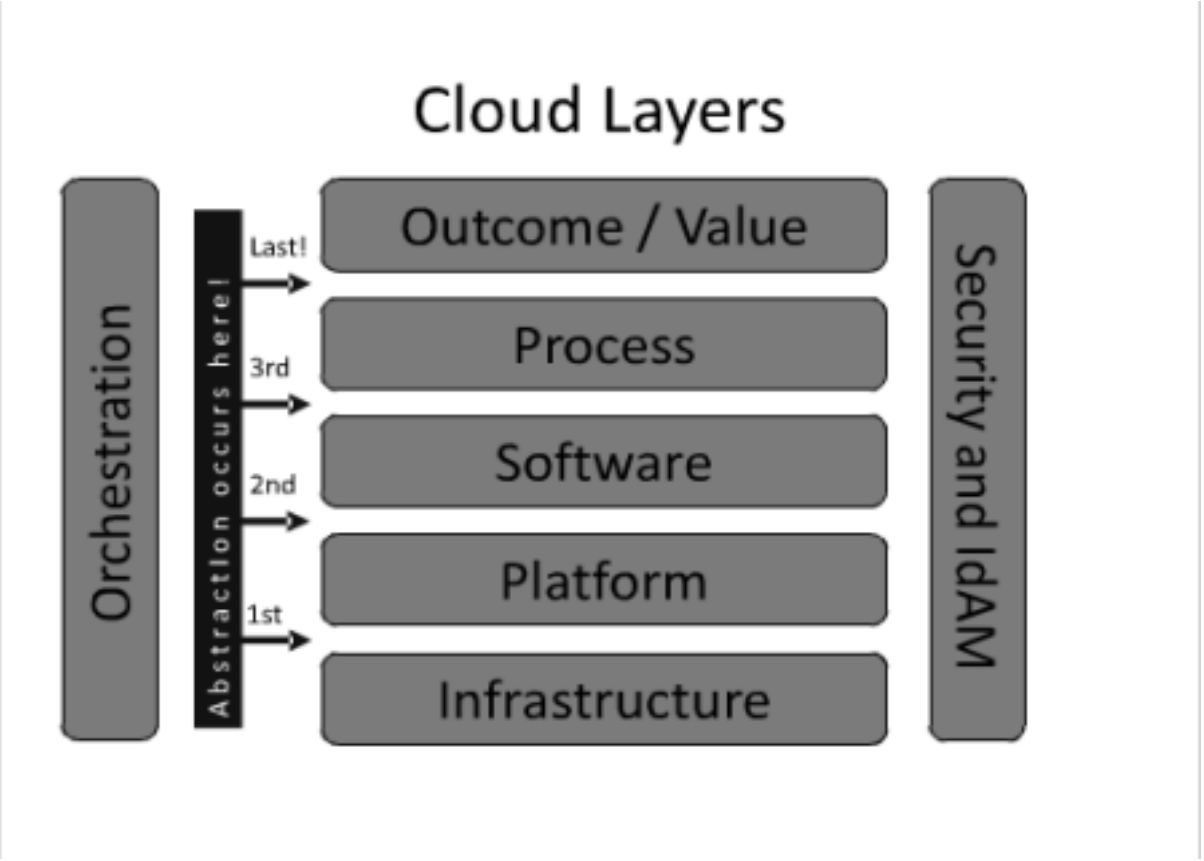

[Fig. 3]

One other element within Confidentiality is the ability to destroy data. In a cloud, that we do not own, and on storage media that we do not control, there is high –probability that the same media be used for other purposes. These storage buckets are dynamic and the service /platform/ application provider might allocate them to other users. This sharing, and in many cases, repeated sharing, of storage media leads to the need for *assured destruction*. We must follow a strict regime that states how long is data to be kept, when and by whom destroyed, and how such destruction is verified. If we go in further detail the question of confidentiality become even more complicated. Also given problem is applicable to both storage and computation units of Cloud Computing. **Integrity** is important factor as well. Because for huge data user must be assured that whatever calculation is done by the cloud is done correctly without any minute errors. Also there should be some procedure that can assure the client that whatever data that will be stored on the file servers that will be stored without tempering any of the data. It will be in the same form and processed it without any assumption about the data. So Integrity requires two questions to be answered those are if data that is being computed is the original data and computation done on the data is error free and produces no harm effects on data or cloud models. **Availability** is most important concern for the users. They must be aware that what the availability ratio of the cloud provider is because availability of their product depends on the availability of the cloud. This is by far the most challenging issue for clouds. User has to be sure that how much of his data is available in case of corruption of existing data or what is the availability of the resources in cloud they are planning to buy. Because if it has no established recovery model or security threat solutions, then economic graph for that product will increase.

## III. Computational Security

**Information security** means protecting information and information systems from unauthorized access, use, disclosure, disruption, modification, perusal, inspection, recording or destruction. For many organizations, security of information is the most critical risk. This may be driven by a need to protect intellectual property, trade secrets, personally identifiable information, or other sensitive information. Making that sensitive information available on the Internet requires a significant investment in security controls and monitoring of access to the content and the pathways to the information. The logging and auditing controls provided by some vendors are not yet as robust as the logging provided within enterprises and enterprise applications. The challenge here is to ensure that, post incident, the organization has visibility to anyone who had access to the document and what might have been done to the document (edit, download, change access, etc.). Governments, military, corporations, financial institutions, hospitals, and private businesses amass a great deal of confidential information about their employees, customers, products, research, and financial status. Most of this information is now collected, processed and stored on electronic computers and transmitted across networks to other computers. Should confidential information about a business' customers or finances or new product line fall into the hands of a competitor, such a breach of security could lead to lost business, law suits or even bankruptcy of the business. Protecting confidential information is a business requirement, and in many cases also an ethical and legal requirement. While these concerns may not be absolute barriers to moving data storage and applications to the cloud environment, clearly they are significant obstacles that will require an enterprise to carefully examine its contractual obligations, risk profile, security infrastructure and oversight ability. An enterprise should be prepared to present the vendor with detailed security and legal requirements applicable to their business needs and the nature of the information being stored or transacted.

Some of the issues while processing information on the cloud are presented below.

A. Abuse and Nefarious use of Cloud Computing
B. Resource Exhaustion
C. Malicious Insider
D. Insecure Interfaces and APIs
E. Account or Service Hijacking

### *A. Abuse and Nefarious use of Cloud Computing*

Cloud providers offer their customers the illusion of unlimited compute, network, and storage capacity — often coupled with a 'frictionless' registration process where anyone with a valid credit card can register and immediately begin using cloud services. Some providers even offer free limited trial periods. By abusing the relative anonymity behind these registration and usage models, spammers, malicious code authors, and other criminals have been able to conduct their activities with relative impunity. The providers have traditionally suffered most from this kind of attacks; Future areas of concern include password and key cracking, DDOS[3], launching dynamic attack points, hosting malicious data, botnet command and control[4], building rainbow tables[5], and CAPTCHA solving farms[6].

*Examples*

Cloud providers have experienced attacks like the Zeus botnet[7], InfoStealer trojan horses and downloads for Microsoft Office and Adobe PDF exploits. Additionally, botnets have used cloud servers for command and control functions. Spam continues to be a problem — as a defensive measure, entire blocks of infected network addresses have been publicly blacklist.

*Remediation*

- Stricter initial registration and validation processes.
- Enhanced credit card fraud monitoring and coordination.
- Comprehensive introspection of customer network traffic.
- Monitoring public blacklists for one's own network blocks.

*Impact*

Criminals continue to leverage new technologies to improve their reach, avoid detection, and improve the effectiveness of their activities. Cloud Computing providers are actively being targeted, partially because their relatively weak registration systems facilitate anonymity, and providers' fraud detection capabilities are limited.

### *B. Resource Exhaustion*

Resource Exhaustion happens when the cloud management does not properly restrict the size or amount of resources that are requested or influenced by an actor, which can be used to consume more resources than intended.

Limited resources include memory, file system storage, database connection pool entries, or CPU. If an attacker can trigger the allocation of these limited resources, but the number or size of the resources is not controlled, then the attacker could cause a denial of service that consumes all available resources. This would prevent valid users from accessing the software, and it could potentially have an impact on the surrounding environment. For example, a memory exhaustion attack against an application could slow down the application as well as its host operating system.

Resource exhaustion problems have at least two common causes:

1. Error conditions and other exceptional circumstances

2. Confusion over which part of the program is responsible for releasing the resource

*Consequences*

- The most common result of resource exhaustion is denial of service. The software may slow down, crash due to unhandled errors, or lock out legitimate users.
- In some cases it may be possible to force the software to "fail open" in the event of resource exhaustion. The state of the software – and possibly the security functionality - may then be compromised.

*Detection Methods*

*Automated Static Analysis*

Automated static analysis [8] typically has limited utility in recognizing resource exhaustion problems, except for program-independent system resources such as files, sockets, and processes. For system resources, automated static analysis may be able to detect circumstances in which resources are not released after they have expired. Automated analysis of configuration files may be able to detect settings that do not specify a maximum value.

Automated static analysis tools will not be appropriate for detecting exhaustion of custom resources, such as an intended security policy in which a bulletin board user is only allowed to make a limited number of posts per day.

Effectiveness: Limited

*Automated Dynamic Analysis*

Certain automated dynamic analysis techniques [8] may be effective in spotting resource exhaustion problems, especially with resources such as processes, memory, and connections. The technique may involve generating a large number of requests to the software within a short time frame.

Effectiveness: Moderate

*Fuzzing*

While fuzzing [9] is typically geared toward finding low-level implementation bugs, it can inadvertently find resource exhaustion problems. This can occur when the fuzzer generates a large number of test cases but does not restart the targeted software in between test cases. If an individual test case produces a crash, but it does not do so reliably, then an inability to handle resource exhaustion may be the cause.

Effectiveness: Opportunistic

*Example*

This code allocates a socket and forks each time it receives a new connection.

```
sock=socket(AF_INET, SOCK_STREAM, 0);
while (1)
{
newsock=accept(sock, ...);
printf("A connection has been accepted\n");
pid = fork();
}
```

The program does not track how many connections have been made, and it does not limit the number of connections. Because forking is a relatively expensive operation, an attacker would be able to cause the system to run out of CPU, processes, or memory by making a large number of connections. Alternatively, an attacker could consume all available connections, preventing others from accessing the system remotely.

### C. Malicious Insider

The threat of a malicious insider is well-known to most organizations. This threat is amplified for consumers of cloud services by the convergence of IT services and customers under a single management domain, combined with a general lack of transparency into provider process and procedure. For example, a provider may not reveal how it grants employees access to physical and virtual assets, how it monitors these employees, or how it analyzes and reports on policy compliance. To complicate matters, there is often little or no visibility into the hiring standards and practices for cloud employees. This kind of situation clearly creates an attractive opportunity for an adversary — ranging from the hobbyist hacker, to organized crime, to corporate espionage, or even nation-state sponsored intrusion. The level of access granted could enable such an adversary to harvest confidential data or gain complete control over the cloud services with little or no risk of detection.

*Remediation*

- Enforce strict supply chain management and conduct a comprehensive supplier assessment.
- Specify human resource requirements as part of legal contracts.
- Require transparency into overall information security and management practices, as well as compliance reporting.
- Determine security breach notification processes.

*Impact*

The impact that malicious insiders can have on an organization is considerable, given their level of access and ability to infiltrate organizations and assets. Brand damage, financial impact, and productivity losses are just some of the ways a malicious insider can affect an operation. As organizations adopt cloud services, the human element takes on an even more profound importance. It is critical therefore that consumers of cloud services understand what providers are doing to detect and defend against the malicious insider

### D. Insecure Interfaces and APIs

Cloud computing providers expose a set of software interfaces or APIs that customers use to manage and interact with cloud services. Provisioning, management, orchestration, and monitoring are all performed using these interfaces. The security and availability of general cloud services is dependent upon the security of these basic APIs. From authentication and access control to encryption and activity monitoring, these interfaces must be designed to protect against both accidental and malicious attempts to circumvent policy. Furthermore, organizations and third parties often build upon these interfaces to offer value-added services to their customers. This introduces the complexity of the new layered API; it also increases risk, as organizations may be required to relinquish their credentials to third- parties in order to enable their agency.

*Examples*

Anonymous access and/or reusable tokens or passwords, clear-text authentication or transmission of content, inflexible access controls or improper authorizations, limited monitoring and logging capabilities, unknown service or API dependencies.

*Remediation*

- Analyze the security model of cloud provider interfaces.
- Ensure strong authentication and access controls are implemented in concert with encrypted transmission.
- Understand the dependency chain associated with the API.

*Impact*

While most providers strive to ensure security is well integrated into their service models, it is critical for consumers of those services to understand the security implications associated with the usage, management, orchestration and monitoring of cloud services. Reliance on a weak set of interfaces and APIs exposes organizations to a variety of security issues related to confidentiality, integrity, availability and accountability.

### E. *Account or Service Hijacking:*

Account or service hijacking is not new. Attack methods such as phishing [10], fraud, and exploitation of software vulnerabilities still achieve results. Credentials and passwords are often reused, which amplifies the impact of such attacks. Cloud solutions add a new threat to the landscape. If an attacker gains access to your credentials, they can eavesdrop on your activities and transactions, manipulate data, return falsified information, and redirect your clients to illegitimate sites. Your account or service instances may become a new base for the attacker. From here, they may leverage the power of your reputation to launch subsequent attacks.

*Remediation*

- Prohibit the sharing of account credentials between users and services.
- Leverage strong two-factor authentication techniques where possible.
- Employ proactive monitoring to detect unauthorized activity.
- Understand cloud provider security policies and SLAs.

*Impact*

Account and service hijacking, usually with stolen credentials, remains a top threat. With stolen credentials, attackers can often access critical areas of deployed cloud computing services, allowing them to compromise the confidentiality, integrity and availability of those services. Organizations should be aware of these techniques as well as common defense in depth protection strategies to contain the damage (and possible litigation) resulting from a breach.

## IV. Storage Security

Many experts in government and commerce still consider the greatest barrier to adoption of cloud services to be concerns about information security and privacy. While these risks exist across the entire cloud ecosystem, every cloud customer retains responsibility for assessing and understanding the value and sensitivity of the data they may choose to move to the cloud. As the owners of that information, cloud customers also remain accountable for decisions regarding the protection of that data wherever it may be stored. Organizations considering moving services to the cloud should keep these information security challenges in mind as they determine cloud adoption strategies:

- A growing interdependence amongst public and private sector entities and the people they serve continues to develop as government, industry, and commercial groups work to establish more widely accepted definitions of cloud computing. While those definitions and the associated standards continue to be created, one cloud requirement is clear—that platform services and hosted applications be secure and available.
- The cloud—however it is defined—is a dynamic hosting environment in which technologies and business models continue to evolve. This continuous change is a security challenge that cloud providers must address through an effective and dynamic security program.
- Sophisticated malicious attempts aimed at obtaining identities or blocking access to sensitive business data threaten to undermine the willingness of organizations to adopt cloud services. Cloud providers must prove that they have put into place and constantly evaluate the effectiveness of the technologies, controls, and processes used to mitigate such disruptions.
- In addition to these challenges, cloud providers must also address the myriad requirements related to delivering services globally online including those coming from governments, legal rulings, and industry standards.

In short, cloud service providers need to manage information security risks in a way that engenders trust with their customers—the government organizations or businesses that do provide such services to end users, as well as directly with end users.
Some of the issues while processing information on the cloud are presented below.

A. Shared Technology Issues
B. Data loss and Leakage
C. Insecure and Ineffective deletion of data

### A. *Shared Technology Issues:*

Cloud vendors deliver their services in a scalable way by sharing infrastructure. Often, the underlying components that make up this infrastructure (e.g., CPU caches, GPUs, etc.) were not designed to offer strong isolation properties for a multi-tenant architecture. To address this gap, a virtualization hypervisor [11] mediates access between guest operating systems and the physical compute resources. Still, even hypervisors have exhibited flaws that have enabled guest operating systems to gain inappropriate levels of control or influence on the underlying platform. A defense in depth strategy is recommended, and should include compute, storage, and network security enforcement and monitoring. Strong compartmentalization should be employed to ensure that individual customers do not impact the operations of other tenants running on the same cloud provider. Customers should not have access to any other tenant's actual or residual data, network traffic, etc.

*Examples*

- Joanna Rutkowska's Red [12] and Blue Pill [13] exploits
- Kortchinksy's CloudBurst presentations. [14]

*Remediation*

- Implement security best practices for installation/configuration.
- Monitor environment for unauthorized changes/activity.
- Promote strong authentication and access control for administrative access and operations.

- Enforce service level agreements for patching and vulnerability remediation.
- Conduct vulnerability scanning and configuration audits.

*Impact*
Attacks have surfaced in recent years that target the shared technology inside Cloud Computing environments. Disk partitions, CPU caches, GPUs, and other shared elements were neverdesigned for strong compartmentalization. As a result, attackers focus on how to impact the operations of other cloud customers, and how to gain unauthorized access to data.

### B. *Data Loss and Leakage:*

There are many ways to compromise data. Deletion or alteration of records without a backup of the original content is an obvious example. Unlinking a record from a larger context may render it unrecoverable, as can storage on unreliable media. Loss of an encoding key may result in effective destruction. Finally, unauthorized parties must be prevented from gaining access to sensitive data. The threat of data compromise increases in the cloud, due to the number of and interactions between risks and challenges which are either unique to cloud, or more dangerous because of the architectural or operational characteristics of the cloud environment.
*Examples*
Insufficient authentication, authorization, and audit controls; inconsistent use of encryption and software keys; operational failures; persistence and remanence challenges: disposal challenges; risk of association; jurisdiction and political issues; data center reliability; and disaster recovery.

*Remediation*
- Implement strong API access control.
- Encrypt and protect integrity of data in transit.
- Analyzes data protection at both design and run time.
- Implement strong key generation, storage and management and destruction practices.
- Contractually demand providers wipe persistent media before it is released into the pool.
- Contractually specify provider backup and retention strategies.

*Impact*
Data loss or leakage can have a devastating impact on a business. Beyond the damage to one's brand and reputation, a loss could significantly impact employee, partner, and customer morale and trust. Loss of core intellectual property could have competitive and financial implications. Worse still, depending upon the data that is lost or leaked, there might be compliance violations and legal ramifications.

*Good Standard for Data Security*
Open PGP [15] is considered as a better standard for data security. Open PGP combines symmetric and asymmetric encryption schemes to form a security model that not only protects the data but does so in a way that is practical and does not compromise the performance of the system. Symmetric encryption, where the same key is used to encrypt and decrypt, tends to be fast at encrypting lots of data. The Advanced Encryption Standard AES – 256 is a symmetric encryption scheme used by the U.S. government.

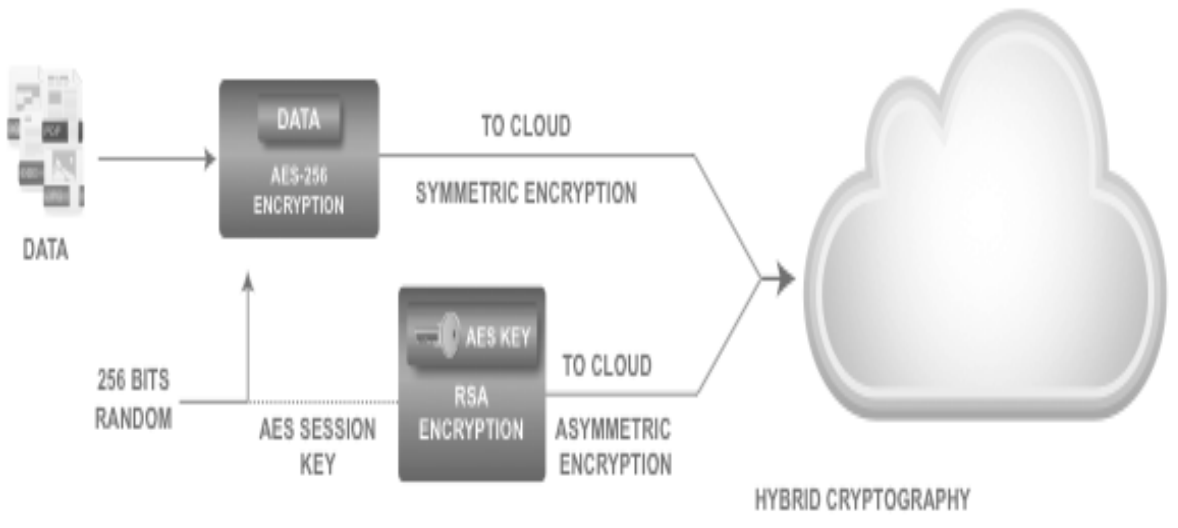

*OpenPGP hybrid encryption to the cloud*

The 256 indicates the size of the key in bits. Open PGP uses symmetric encryption like AES-256 to encrypt data and asymmetric encryption like RSA (Rivest-Shamir-Aldeman) to encrypt the keys used by AES-256. Asymmetric encryption simplifies key management, but is generally slower than symmetric encryption. This hybrid approach using the fast symmetric encryption to encrypt data and the slower asymmetric encryption only to encrypt the (comparatively small) keys allows data to be encrypted efficiently and a high level of granularity. Every data packet in the cloud can be protected separately with its own symmetric key and those keys can be managed together through their combined asymmetric key. This allows a practical level of control and granularity of keys and encrypted objects. The asymmetric keys can be maintained by the user in a key ring that becomes the single point of access control to the whole system.

### C. *Insecure or Ineffective deletion of data*

Whenever a provider is changed, resources are scaled down, physical hardware is reallocated, etc, data may be available beyond the lifetime specified in the security policy. It may be impossible to carry out the procedures specified by the security policy, since full data deletion is only possible by destroying a disk, which also stores data from other clients. When a request to delete a cloud resource is made, this may not result in true wiping of the data (as with most operating systems). Where true data wiping is required, special procedures must be followed and this may not be supported by the standard API (or at all).

If effective encryption is used then the level of risk may be considered to be lower.

*Remediation*
Good encryption strategies
Good Timely deletion Strategies

*Impact*
Personal sensitive data and credentials are affected.

## V. Network security

Since cloud computing uses the Internet as the communication media for providing different computing services like SaaS,PaaS, IaaS, it is vulnerable to various network security threats. This section explains various network security threats that could occur on the cloud and the possible ways of prevention/mitigation of those attacks.

The following are some of the network security threats that can cause damage on the cloud computing system.

A. Flooding attacks such as Dos and DDos
B. Data Interception attacks
C. Management Interface attacks
D. Cloud Malware attacks
E. Metadata spoofing attacks

### A. *Flooding Attacks*

A major aspect of Cloud Computing consists in outsourcing basic operational tasks to a Cloud system provider [16]. Among these basic tasks, one of the most important ones is server hardware maintenance. Thus, instead of operating an own, internal data center, the paradigm of Cloud Computing enables companies (users) to *rent* server hardware on demand (IaaS). This approach provides valuable economic benefits when it comes to dynamics in server load, as for instance day-and-night cycles can be attenuated by having the data traffic of different time zones operated by the same servers. Thus, instead of buying sufficient server hardware for the high workload times, Cloud Computing enables a dynamic adaptation of hardware requirements to the actual workload occurring.
Technically, this achievement can be realized by using virtual machines deployed on arbitrary data center servers of the Cloud system. If a company's demand on computational power rises, it simply is provided with more instances of virtual machines for its services. Under security considerations, this architecture has a serious drawback. Though the feature of providing more computational power on demand is appreciated in the case of valid users, it poses severe troubles in the presence of an attacker. The corresponding threat is that of *flooding attacks*, which basically consist in an attacker sending a huge amount of nonsense requests to a certain service. As each of these requests has to be processed by the service implementation in order to determine its invalidity, this causes a certain amount of workload per attack request, which—in the case of a flood of requests—usually would cause a Denial of Service to the server hardware [16]. In the specific case of Cloud Computing systems, the impact of such a flooding attack is expected to be amplified drastically.

*Direct Denial of Service*

When the Cloud Computing operating system notices the high workload on the flooded service, it will start to provide more computational power (more VMs) more service instances...) to cope with the additional workload. Thus, the server hardware boundaries for maximum workload to process do no longer hold. In that sense, the Cloud system is trying to work *against* the attacker (by providing more computational power), but actually—to some extent—even *supports* the attacker by enabling him to do most possible damage on a service's availability, starting from a single flooding attack entry point. Thus, the attacker does not have to flood all *n* servers that provide a certain service in target, but merely can flood a single, Cloud-based address in order to perform a full loss of availability on the intended service [16].

*Indirect Denial of Service Attacks*

Depending on the computational power in control of the attacker, a side effect of the direct flooding attack on a Cloud service potentially consists in that other services provided on the same hardware servers may suffer from the workload caused by the flooding. Thus, if a service instance happens to run on the same server with another, flooded service instance, this may affect its own availability as well. Once the server's hardware resources are completely exhausted by processing the flooding attack requests, obviously also the other service instances on the same hardware machine are no longer able to perform their intended tasks. Thus, the Denial of Service of the targeted service instances are likely to cause a Denial of Service on all other services deployed to the same server hardware as well.
Depending on the level of sophistication of the Cloud system, this side-effect may worsen if the Cloud system notices the lack of availability, and tries to "evacuate" the affected service instances to other servers. This results in additional workload for those other servers, and thus the flooding attack "jumps over" to another service type, and spreads throughout the whole computing Cloud. In the worst case, an adversary manages to utilize another (or the very same) Cloud Computing system for hosting his flooding attack application. In that case, the race in power would play both Cloud systems off against each other; each Cloud would provide more and more computational resources for creating, respectively fending, the flood, until one of them eventually reaches full loss of availability [16].
*Examples*

The following is one of the incidents of a Dos attack on Amazon cloud Posted in Enterprise Security, 5th October 2009 15:32 GMT

***"DDoS attack rains down on Amazon cloud"***

*Web-based code hosting service Bitbucket experienced more than 19 hours of downtime over the weekend after an apparent DDoS attack (flooding of millions of UDP packets) on the sky-high compute infrastructure it rents from Amazon.com.*

*Remediation*

- Usage of load balancers to mitigate the incoming aggregated traffic by routing the requests to different servers.
- Anycast networking concept wherein the same content is served from different physical and geographical servers.

- Blackholing - Traffic to victim is redirected to a black hole(null interface, invalid server etc)
- Sinkholing using in-depth packet inspection

*Remediation strategies used by Cloud Providers:-*

*Microsoft:-*

- Microsoft applies several layers of security as appropriate to data center devices and network connections [17]
- Specialized hardware such as load balancers, firewalls, and intrusion prevention devices, is in place to manage volume-based denial of service (DoS) attacks [17].
- Through network hardware, Microsoft uses application gateway functions to perform deep packet inspection and take actions such as sending alerts based on—or blocking—suspicious network traffic[17].

*Amazon:-*

- Uses standard DDoS mitigation techniques such as sync cookies and connection limiting [18].
- Amazon maintains internal bandwidth which exceeds its provider-supplied Internet bandwidth.

### B. *Cloud Malware Injection Attack*

A first considerable attack attempt aims at injecting a malicious service implementation or virtual machine into the Cloud system. Such kind of **Cloud malware** could serve any particular purpose the adversary is interested in, ranging from eavesdropping via subtle data modifications to full functionality changes or blockings. This attack requires the adversary to create its own malicious service implementation module (SaaS or PaaS) or virtual machine instance (IaaS), and add it to the Cloud system. Then, the adversary has to trick the Cloud system so that it treats the new service implementation instance as one of the valid instances for the particular service attacked by the adversary. If this succeeds, the Cloud system automatically redirects valid user requests to the malicious service implementation, and the adversary's code is executed [16].

*Remediation*

A promising countermeasure approach to this threat consists in the Cloud system performing a service instance integrity check prior to using a service instance for incoming requests. This can e.g. be done by storing a hash value on the original service instance's image file and comparing this value with the hash values of all new service instance images. Thus, an attacker would be required to trick that hash value comparison in order to inject his malicious instances into the Cloud system [16]. Another approach to counter malware attack is to periodically scan the cloud systems for any suspected application such as worm/Trojan/malware etc.

*Remediation strategies used by Cloud Providers:-*

*Amazon:-*

Amazon uses HackAlert™ [20], a malware monitoring and detection software delivered as SaaS to protect the customer websites from cloud malware attack. HackAlert™ connects to the monitored website over a standard HTTP connection and captures all responses in deliberately unsecured "Honey Clients" located at Armorize data centers worldwide. All website responses are analyzed for the presence of both active malware content and suspicious links (to external sites not currently distributing malware). This distinction greatly reduces the amount of false positives.

*Impact*

Any malware attack could destroy the intellectual property of the cloud provider as well as the customers as their confidential data could be kept in the cloud system. Usually a malware attack attempts to retrieve the user credential information and use the same to retrieve critical information from the system. This type of attack could degrade the reputation and trust of the cloud provider.

*Examples*

***"MALWARE ATTACK USES CHINA WORLD EXPO GUISE" [19]***

Posted by Owen Fletcher March 25, 2010 06:12 AM ET

A malware attack dressed up as an e-mail from organizers of the upcoming Shanghai World Expo targeted at least three foreign journalists in China, in the latest sign of increasingly sophisticated cyber attacks from the country.
The e-mail appeared to be sent from the inbox of the Expo news office, but it was not sent by the Expo and may be targeting journalists who signed up to cover the event

### C. *Data Interception Attack*

Cloud computing, being a distributed architecture, implies more data in transit than traditional infrastructures. For example, data must be transferred in order to synchronise multiple distributed machine images, images distributed across multiple physical machines, between cloud infrastructure and remote web clients, etc. Furthermore, most use of data-centre hosting is implemented using a secure VPN-like connection environment, a practice not always followed in the cloud context. Sniffing, spoofing, man-in–the-middle attacks, side channel and replay attacks should be considered as possible threat sources. Moreover, in some cases the Cloud Provider does not offer a confidentiality or non-disclosure clause or these clauses are not sufficient to guarantee respect for the protection of the customer's secret information and 'know-how' that will circulate in the 'cloud' [21].

*Types of Data Interception Attacks*

*Sniffing*
This attack involves sniffing and manipulating packets flowing through the cloud network or between web browser and the cloud system.

*Spoofing*
This kind of interception is done by sending illegitimate connection requests and messages from invalid sources. The scatter effect produced are utilized to produce further attacks on the cloud.

*Man-In-The-Middle (MITM) attacks*
This kind of attack involves interception of traffic by being in the middle of the traffic flowing between the cloud and the intended recipient. Spoofed data is sent to both the endpoints.

*Side Channel Attack*
This kind of attack involves using timing information, power consumption, electromagnetic leaks or even sound to break the system.

*Possible Threat sources*
AAA (Authentication, Authorization, Accounting) Vulnerability

A poor system for authentication, authorization and accounting, could facilitate unauthorized access to resources, privileges escalation, impossibility of tracking the misuse of resources and security incidents in general, etc, through:

- Insecure storage of cloud access credentials by customer
- Insufficient roles available
- Credentials stored on a transitory machine.

Furthermore, the cloud makes password based authentication attacks (trend of fraudster using a Trojan to steal corporate passwords) much more impactful since corporate applications are now exposed on the Internet. Therefore password-based authentication will become insufficient and a need for stronger or two-factor authentication for accessing cloud resources will be necessary [21].

*Communication Encryption vulnerabilities*
These vulnerabilities concern the possibility of reading data in transit via, for example, MITM attacks, poor authentication, acceptance of self-signed certificates, etc [21].

*Weak Encryption of archives and data in Transit*

Failure to encrypt data in transit, data held in archives and databases, un-mounted virtual machine images, forensic images and data, sensitive logs and other data at rest puts the data at risk. Of course the costs of implementing key management and processing costs must be taking account and set against the business risk introduced [21].

*Remediation*

The strategy to counter date interception attacks is to
- Have a strong AAA system which does not expose any vulnerability for unauthorized access/unclear role definitions.
- Have a strong encryption scheme for the data and control traffic between the cloud systems as well as between the customer and the cloud provider.

### *D. Management Interface Attack*

The customer management interfaces of public cloud providers are Internet accessible and mediate access to larger sets of resources (than traditional hosting providers) and therefore pose an increased risk especially when combined with remote access and web browser vulnerabilities. This includes customer interfaces controlling a number of virtual machines and, most importantly, Cloud Provider interfaces controlling the operation of the overall cloud system. Of course, this risk may be mitigated by more investment in security by providers [21].

*Threat Sources*

One of the sources for the management interface attack is the AAA vulnerability. Lack of or inefficient challenge response system during the authentication through remote clients could cause attack on the management interfaces.
Another possible source of this attack could be misconfiguration of specific key parameters of the cloud system. This could be due to:
- Inadequate application of security baseline
- Invalid or incorrect implementation of hardening procedures
- Human error and untrained administrator

Misconfiguration or a known OS or System vulnerability could also cause a management interface attack. For example conflicting patching procedures used between the customer and the cloud provider could result in misconfiguration of the cloud system.

*Remediation*
Management interfaces should be exposed in the form of a secure channel. Instead of the password based authentication, it should use two-factor authentication. Periodic and efficient OS and hardware hardening procedures should be followed on the cloud system.

## VI. CONCLUSION

Cloud computing is the next big wave in computing. It has many benefits, such as better hardware management, since all the computers are the same and run the same hardware. It also provides for better and easier management of data security, since all the data is located on a central server, so administrators can control who has and doesn't have access to the files. It is widely accepted today because of its economic benefits.

There are some down sides as well to cloud computing. Out of those down falls one of the major factors is security. User will have to evaluate the security model that is been used by Cloud Provider makes lot of impact on taking the decision of the selecting the cloud provider. Also for Cloud Computing there is more number of threats than compare to security of single PC because clouds have many elements than single PC.